\documentclass[preprint,12pt]{elsarticle}

\usepackage{amssymb}
\usepackage{setspace}
\usepackage{amsfonts}

\journal{Physica A, (2010) in press}

\begin{document}

\begin{frontmatter}

\title{The significant digit law in statistical physics}

\author{Lijing Shao}
\author{Bo-Qiang Ma}
\ead{mabq@phy.pku.edu.cn}

\address{School of Physics and State Key Laboratory of Nuclear
Physics and Technology, \\
Peking University, Beijing 100871, China}

\begin{abstract}

The occurrence of the nonzero leftmost digit, i.e., 1, 2, ..., 9, of
numbers from many real world sources is not uniformly distributed as
one might naively expect, but instead, the nature favors smaller
ones according to a logarithmic distribution, named Benford's law.
We investigate three kinds of widely used physical statistics, i.e.,
the Boltzmann-Gibbs (BG) distribution, the Fermi-Dirac (FD)
distribution, and the Bose-Einstein (BE) distribution, and find that
the BG and FD distributions both fluctuate slightly in a periodic
manner around the Benford distribution with respect to the
temperature of the system, while the BE distribution conforms to it
exactly whatever the temperature is. Thus the Benford's law seems to
present a general pattern for physical statistics and might be even
more fundamental and profound in nature. Furthermore, various
elegant properties of Benford's law, especially the mantissa
distribution of data sets, are discussed.

\end{abstract}


\end{frontmatter}

\clearpage

\section{Introduction}

One may simply presume that occurrence of the first digit of any
randomly chosen data set is approximately uniformly distributed, but
that is not the very case in real world. In 1881, Newcomb~\cite{n81}
noticed that the preceding pages of the logarithmic table wear out
faster, thus he hinted at the idea that the first nonzero digit of
many natural numbers favors small values, where the number $1$
appears almost seven times more often than that of the number $9$.
Then independently in 1938, Benford~\cite{b38} investigated a great
number of data sets in various unrelated fields, and found that they
agree with a logarithmic distribution, which now we refer to as
Benford's law after the name of its second discoverer,
\begin{equation}\label{ben}
P_{\rm Ben}(d) = \log_{10} (1 + \frac{1}{d}), \, d=1,2,...,9
\end{equation}
where $P_{\rm Ben}(d)$ is the probability of a number having the first digit $d$.
It is also named the first digit law or the significant
digit law.

The discoveries of various samples from different domains that
comply with Benford's law are accumulating significantly all these
years. Empirically, the areas of lakes, the lengths of rivers, the
arabic numbers on the front page of a newspaper~\cite{b38}, physical
constants~\cite{bk91}, the stock market indices~\cite{l96}, file
sizes in a personal computer~\cite{tfgs07}, survival
distributions~\cite{lse00}, widths of hadrons~\cite{sm09a}, various
quantities of pulsars~\cite{sm09b}, even some general dynamical
systems~\cite{tbl00,bbh05,b05}, all conform to the peculiar law
well. Nevertheless, there also exist other types of data which do
not obey the law, e.g., lottery and telephone numbers.
Unfortunately, there is no a priori criteria yet to judge which type
a data set belongs to. In practice, the law is already applicable in
distinguishing and ascertaining fraud in taxing and
accounting~\cite{n96,nm97,rr03,gw04}, as well as speeding up
calculation and minimizing expected storage space in computer
science~\cite{bb85,s88,bh07}.

Since its second discovery in 1938, many attempts have been tried to
explain the underlying reason for Benford's law. For theoretical
reviews, see papers written by Raimi~\cite{r69a,r69b,r76,r85} and
Hill~\cite{h95a,h95b,h95c,h98,hs05}. Nowadays, many breakthrough
points have been achieved, though, there still lacks a universally
accepted final answer. In mathematics, Benford's law is the only
digit law that is scale-invariant~\cite{bhm08}, which means that the
law does not depend on any particular choice of units, first
discovered by Pinkham~\cite{p61}. Benford's law is also
base-invariant~\cite{h95a,h95b,h95c}, which means that it is
independent of the base $b$. In the binary system ($b$=2), octal
system ($b$=8), or other base system, the data, as well as in the
decimal system ($b$=10), all fit the general Benford's law,
\begin{equation}\label{benbase}
P_{\rm Ben}(d) = \log_{b}(1+\frac{1}{d}), \, d=1, 2, ..., {b-1}.
\end{equation}
Theoretically, Hill proved that ``scale-invariance
implies base-invariance''~\cite{h95a} and ``base-invariance
implies Benford's law''~\cite{h95b} mathematically in the
framework of probability theory.

It was suggested that ``an interesting open problem is to determine
which common distributions (or mixtures thereof) satisfy Benford's
law''~\cite{h95c}. To track the suggestive idea~\cite{lse00} and
explore more aspects of the peculiar law in general physics, we
study comprehensively three kinds of widely used physical statistics
in this paper, i.e., the Boltzmann-Gibbs (BG) distribution, the
Fermi-Dirac (FD) distribution, and the Bose-Einstein (BE)
distribution. We find that the BG and FD distributions both
fluctuate slightly in a periodic manner around the Benford
distribution with respect to the temperature of the system, while
the BE distribution conforms to it exactly, independent of the
temperature of the system. In some sense, the logarithmic
distribution presents a general feature in statistical physics. Thus
it seems that the significant digit law takes precedence over
physical statistics, and it might be a more fundamental principle
behind the complexity of the nature. There are other sorts of
regularities very interesting and not totally understood
yet~\cite{sz09}. Before entering to the details of three statistics,
we discuss the mantissa distribution firstly in the following
section, as it is a closely relevant and vital issue to Benford's
law.

\section{Probability Density of Mantissa}

\label{mantissa}

The mantissa $m \in (\,-1,~-0.1\,]\cup [\,0.1,~1\,)$ is the
significand part of a floating-point number $x$, defined uniquely as
$x = m \times 10^n$, where $n$ is an integer. In this paper, if not
noted explicitly, we always postulate that the numbers are positive
for succinct statement.

Benford's law was stated as that ``the law of probability of the
occurrence of numbers is such that all mantissae of their logarithms
are equally likely''~\cite{n81}. In other words, if we express the
mantissa \{\,$m$\,\} of a data set into the form \{\,$10^t$\,\}
where $t \in [\,-1,~0\,)$, then \{\,$t$\,\} is uniformly
distributed.

The above statement is sufficient to Benford distribution, however,
it is not indispensable to Eq.~(\ref{ben}) and Eq.~(\ref{benbase}).
Actually, the uniform distribution of the logarithm of mantissa~\cite{h95b} is
equivalent to the $n$-digit Benford's law,
\begin{equation}\label{benndigit}
P_{\rm Ben}(d_1,d_2,...,d_n) = \log_{10} \left [1 + \left( \sum_{i=1}^{n}
d_i \times 10^{n-i} \right)^{-1} \right]
\end{equation}
where $d_n$ is the $n$-th leftmost digit, hence
$d_1 \in \{~1,2,...,9~\}$ and $d_n \in \{~0,1,2,...,9~\}$ for $n \ge 2$.
For instance, the probability of finding a number with the first two
digits $d_1=1$ and $d_2=5$ is $P_{\rm Ben}(d_1=1,d_2=5) = \log_{10}(1+1/15) = 2.80\%$.

Under the assumption that the logarithm of mantissa is uniformly
distributed, it is direct to attain the normalized probability
density of mantissa,
\begin{equation}\label{mpd}
f(m) = \frac{1}{\ln 10} \cdot \frac{1}{m}, \, m \in [\,0.1,~1\,)
\end{equation}
which was also achieved by Lemons in terms of a probabilistic model
of partitioning a conserved quantity~\cite{l86} and Pietronero {\it
et al.} using processes where the time evolution is governed by
multiplicative fluctuations~\cite{pttv01}. We suspect that the
probability density of mantissa, i.e., Eq.~(\ref{mpd}), might be a
crucial insight to understand Benford's law in the framework of
probability theory or statistical analysis. However, the generally
mathematical manipulation of the mapping from real data sets to
mantissa sets is hard to deal with. Difficulties notwithstanding,
mantissa distributions induced by distributions on $\mathbb{R}$ have
been studied extensively~\cite{h95b,bhm08}.

\section{Physical Statistics and Benford's Law}

In statistical physics, the BG distribution, the FD distribution,
and the BE distribution are three kinds of widely used canonical
statistics, and have extensive applications in various domains of
physics. On the other hand, many physical samples are found to
reproduce the Benford distribution empirically. Therefore it is
speculated naturally that these Benford samples might present some
general characteristics originating from physical statistics. It
seems very intriguing to look into the relationship between
canonical statistics and Benford's law.

To relate physical statistics to Benford's law quantitatively, we
refer to the language of probability density. Assuming a measurable
quantity, for example, the energy $E$, has its normalized
probability density $f(E)$,
where $E \in (\,0,~+\infty\,)$, then the probability that the energy
$E$ has its significant digit $d$ equals to
\begin{equation}\label{pdsum}
P(d) = \sum_{n = -\infty}^{\infty} \int_{d \cdot 10^n}^{(d+1) \cdot 10^n}
f(E) {\rm d} E
\end{equation}
which will be utilized frequently in the following analysis.

Throughout the whole paper, we use the energy $E$ of the system as
the measurable quantity, whose first digit distribution is compared
with Benford's law as an example. However, the situation should not
be limited to this special case. Since many quantities are
distributed in the same form as or a similar form to the energy $E$,
the results can be extended without any difficulty. The method can
further be extended easily to analyse other kinds of distributions
as well.

\subsection{The Boltzmann-Gibbs distribution}

The Boltzmann-Gibbs (BG) distribution applies to particles under the
classical circumstances where quantum effects can be ignored. It was
discovered by Boltzmann in 1877~\cite{b77}, and significantly
developed by Gibbs in 1901~\cite{g01}, and since then, it plays a
fundamental role throughout statistical physics as a milestone.

To be more specific, for a thermal bath at a well-defined
temperature $T$, the Boltzmann-Gibbs statistics gives the normalized
probability density,
\begin{equation}\label{bg}
f_{BG}(E) = \beta e^{-\beta E}
\end{equation}
where $\beta = 1/kT$, and $k$ is the Boltzmann constant. Eq.~(\ref{bg})
can be derived directly from Liouville's theorem.

Utilizing Eq.~(\ref{pdsum}),
we can immediately obtain the probability of $E$ with its
first nonzero digit $d$,
\begin{equation}\label{bgd}
P_{BG}(d; \beta) = \sum_{n = -\infty}^{\infty} \left[
e^{-\beta d \cdot 10^n} - e^{-\beta (d+1) \cdot 10^n} \right]
\end{equation}
which is a function of both $\beta$ and $d$. It was easily noticed
that $P_{BG}(d; \beta) > P_{BG}(d+1; \beta)$ after factorizing
the summed term into the form of
$e^{-\beta d \cdot 10^n} (1-e^{-\beta \cdot 10^n})$.
It reproduces the rough property of the first digit law qualitatively thereof.

\begin{figure}
\includegraphics[width=14.5cm]{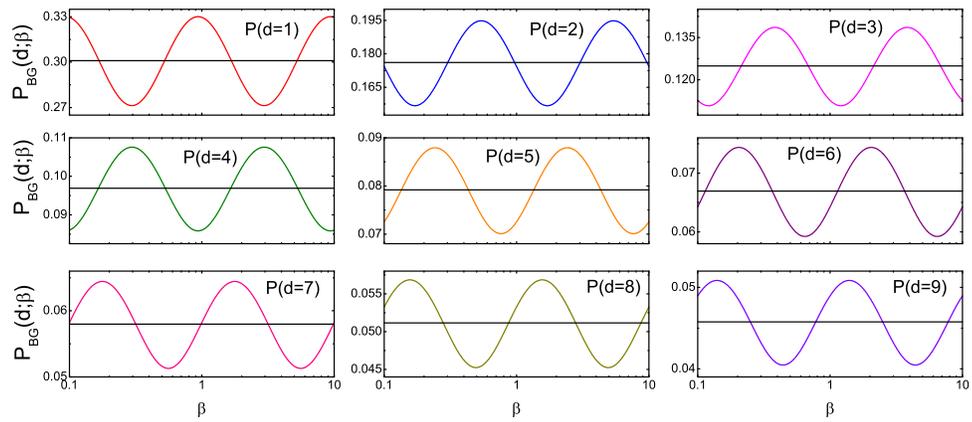}
\caption{The comparisons of the first digit distributions for BG
distribution and Benford's law via the inverse of temperature $\beta
= 1/kT$, where $k$ is the Boltzmann constant. The horizontal line in
each panel presents the value predicted by Benford's law (also see
Figure 1 in Ref.~\cite{el03}).\label{pbg}}
\end{figure}

Further from Eq.~(\ref{bgd}), it is straightforward to check an important fact that
\begin{equation}\label{mbg}
P_{BG}(d; \beta) = P_{BG}(d; 10\beta).
\end{equation}
It was recognized in the exponential random variables by Engel and
Leuenberger~\cite{el03}. This property leads to a periodic function
of $P_{BG}(d; \beta)$ on the $\beta$-logarithmic scale, illustrated
in Fig.~\ref{pbg} (also see Figure 1 in Ref.~\cite{el03}). The
horizontal line in each panel is the value predicted by Benford's
law. It is shown clearly that the first digit distributions of the
BG distribution conform to Benford's law approximately, and
fluctuate around it slightly. The variation is within the bound of
0.03 for number 1~\cite{el03}, and smaller for other digits. The
maximum deviations and the maximum relative deviations are listed in
Table~\ref{terr}.

Furthermore, a new function $P^\star_{BG}(d; \alpha)$ is defined as
$P^\star_{BG}(d; \alpha) = P_{BG}(d; \beta = 10^\alpha)$, which is a
1-periodic function with respect to $\alpha$, i.e., $P^\star_{BG}(d;
\alpha) = P^\star_{BG}(d; \alpha+1)$. Then it was proved that the
Fourier coefficient $c_0$ of the Fourier series $P^\star_{BG}(d;
\alpha) = \sum_n c_n e^{i \cdot 2n \pi \alpha}$ equals exactly to
$\log_{10} (1 + 1/d)$, thus the integral mean of $P^\star_{BG}(d;
\alpha)$ follows Benford's law~\cite{el03}
\begin{equation}
\int_0^1 P^\star_{BG}(d; \alpha) {\rm d} \alpha = \log_{10} (1 + \frac{1}{d})
\equiv  P_{\rm Ben}(d).
\end{equation}
However, the underlying assumption that $\alpha = \log_{10} \beta$
is uniformly distributed appears indiscretionary from the standpoint
of statistical physics.

\subsection{The Fermi-Dirac distribution}

Empirically, not only data sets rooting in macrocosmic systems, but
also those in microcosmic systems obey Benford's law, e.g., widths
of hadrons~\cite{sm09a}, half-lives of
$\alpha$-decays~\cite{bmp93,nr08}, and the strengths of
electric-dipolar lines in transition arrays of complex atomic
spectra~\cite{p08}. Therefore, it is also necessary to study the
first digit distributions of quantum statistics.

The Fermi-Dirac (FD) distribution was proposed by Fermi in 1926 for
electrons~\cite{f26}, and its relation to quantum mechanics was
elucidated by Dirac later~\cite{d26}. It is the statistics obeyed by
particles that are described by antisymmetrical wave function with
half-integral spin, where Pauli principle applies, and the
normalized probability density is
\begin{equation}\label{fd}
f_{FD}(E) = \frac{\beta}{\ln 2} \frac{1}{e^{\beta E}+1}.
\end{equation}

\begin{figure}
\includegraphics[width=14.5cm]{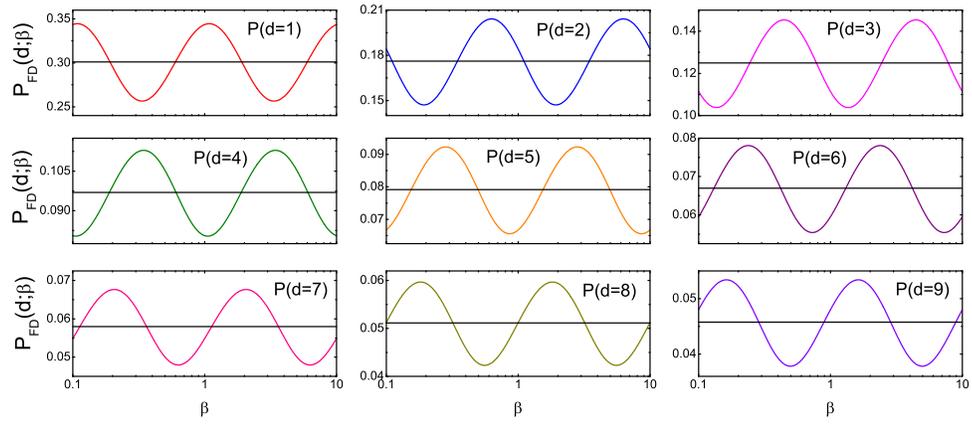}
\caption{The comparisons of the first digit distributions for FD
distribution and Benford's law via the inverse of temperature $\beta
= 1/kT$, where $k$ is the Boltzmann constant. The horizontal line in
each panel presents the value predicted by Benford's
law.\label{pfd}}
\end{figure}

Utilizing Eq.~(\ref{pdsum}), it is straightforward to obtain the
desired probability,
\begin{equation}\label{fdpd}
P_{FD}(d; \beta) = \sum_{n = -\infty}^{\infty} \frac{1}{\ln 2}
\left\{ \beta \cdot 10^n + \ln \left[ \frac{e^{\beta d \cdot
10^n}+1}{e^{\beta (d+1) \cdot 10^n}+1} \right] \right\},
\end{equation}
and results are shown in Fig.~\ref{pfd}, where the horizontal line
in each panel presents the value predicted by Benford's law.

The features of the probability functions are astonishingly similar
to the BG case, with the first digit distributions fluctuating
slightly around the Benford value. However, the variations are
somehow larger than the BG distribution, within the bound of 0.045
for number 1, and smaller for other numbers. The maximum deviations
and the maximum relative deviations are listed in Table~\ref{terr}.

In addition, the useful identity $P_{FD}(d; \beta) = P_{FD}(d;
10\beta)$ still holds in the FD statistics, thus the fluctuation
here is periodic in the $\beta$-logarithmic scale too. We claim that
this property should be a general conclusion originating from the
multiplicative appearance of $\beta$ and $d$ in Eq.~(\ref{bg}) and
Eq.~(\ref{fd}).

Similarly, we define $P^\star_{FD}(d; \alpha) = P_{FD}(d; \beta =
10^\alpha)$, which is a 1-periodic function with respect to
$\alpha$. Then the Fourier coefficient $c_0$ of the Fourier series
$P^\star_{FD}(d; \alpha) = \sum_n c_n e^{i \cdot 2n \pi \alpha}$
equals to
\begin{eqnarray}\label{fdc0}
c_0 & = & \int_0^1 P^\star_{FD}(d; \alpha) {\rm d} \alpha \\ \nonumber
& = & \int_0^1 \sum_{n = -\infty}^{\infty} \frac{1}{\ln 2} \left\{
10^{n+\alpha} + \ln \left[
\frac{e^{d \cdot 10^{n+\alpha}}+1}{e^{(d+1) \cdot 10^{n+\alpha}}+1}
\right] \right\} {\rm d} \alpha\\ \nonumber
& = & \int_{-\infty}^{\infty} \frac{1}{\ln 2} \left\{
10^{\alpha} + \ln \left[
\frac{e^{d \cdot 10^{\alpha}}+1}{e^{(d+1) \cdot 10^{\alpha}}+1}
\right] \right\} {\rm d} \alpha\\ \nonumber
& = & \int_0^\infty \frac{1}{\ln 2 \cdot \ln 10} \left\{
u + \ln \left[
\frac{e^{d \cdot u}+1}{e^{(d+1) \cdot u}+1}
\right] \right\} \frac{{\rm d}u}{u}\\ \nonumber
& \equiv & F(d)
\end{eqnarray}
where a transformation $u=10^\alpha$ is adopted in calculation, and
$c_0$ is denoted as a function $F(d)$ versus the variable $d$.

It is checked that the derivative of $F(d)$ is
\begin{equation}
F^\prime(d) = \frac{1}{\ln 10} \left(
\frac{1}{1+d} - \frac{1}{d} \right),
\end{equation}
and the normalization condition is $ \sum_{d=1}^{9} F(d) = 1$,
thus $c_0 \equiv F(d) = \log_{10}(1+1/d)$,
which is the precise form of Benford's law again. It leads to the
conclusion that the integral mean of $P^\star_{FD}(d; \alpha)$ also
follows the significant digit law, i.e.,
\begin{equation}
\int_0^1 P^\star_{FD}(d; \alpha) {\rm d} \alpha = \log_{10} (1 + \frac{1}{d})
\equiv  P_{\rm Ben}(d).
\end{equation}

\subsection{The Bose-Einstein distribution}

The Bose-Einstein (BE) distribution was introduced by Bose in
1924~\cite{b24}, and generalized by Einstein~\cite{e24}. It is the
statistics obeyed by particles that are described by symmetrical
wave function with integral spin. Its well-known probability density
is
\begin{equation}\label{be}
f_{BE}(E) \propto \frac{1}{e^{\beta E}-1}
\end{equation}
which can not be normalized to unity however. The divergence occurs
when $E$ approaches zero, where the behavior of $f_{BE}(E)$ tends to
be $1/E$. The main (and also total) contribution comes from the
divergent region around the singularity $E=0$ where $f_{FD}
\rightarrow 1/E$. Consequently, the first digit distribution
approaches precisely to the $n$-digit Benford's law according to
Eq.~(\ref{mpd}) and discussions in Section~\ref{mantissa} (also see
Ref.~\cite{h95b}). Thus we claim that the BE statistics follows
Benford's law exactly.


\section{Discussion}

\begin{figure}
\includegraphics[width=13.5cm]{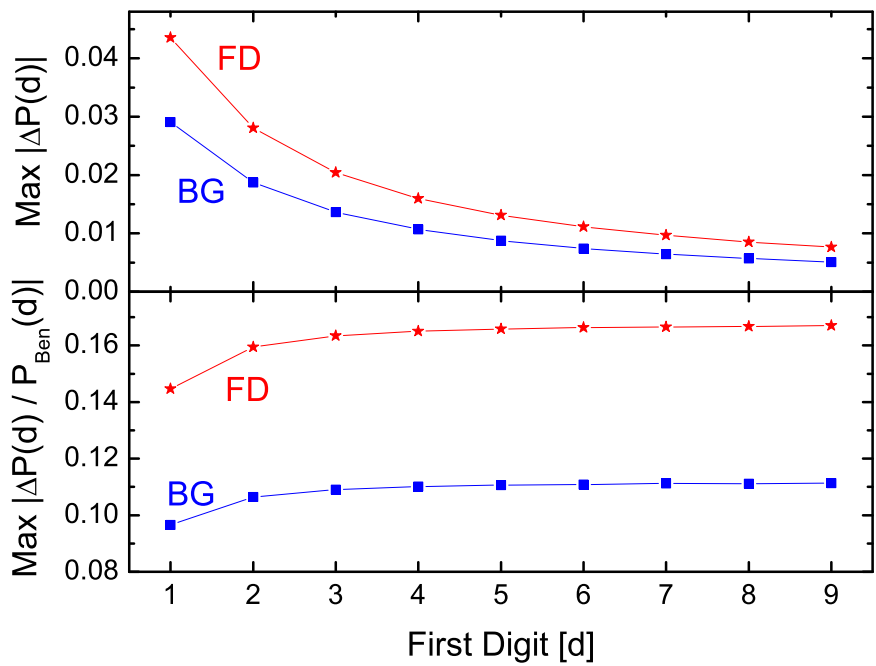}
\caption{{\it Upper panel:} The maximum deviation of the first digit
distributions from Benford's law for the BG distribution (squared)
and the FD distribution (star-shaped); {\it Lower panel:} the
maximum relative deviation of the first digit distributions from
Benford's law for the BG distribution (squared) and the FD
distribution (star-shaped).\label{perr}}
\end{figure}

In Fig.~\ref{perr}, we illustrate graphically the maximum deviation,
Max~$|\Delta P(d)|$, and the maximum relative deviation,
Max~$|\Delta P(d) / P_{\rm Ben}(d)|$, for the BG distribution and
the FD distribution. From the figure, we find that the absolute
value of the maximum deviation decreases monotonously as a function
of $d$, while the maximum relative deviation increases versus $d$.
They both approach to a steady constant individually. The maximally
deviated values of BG statistics are always smaller than those of FD
statistics.

\begin{figure}
\includegraphics[width=13.5cm]{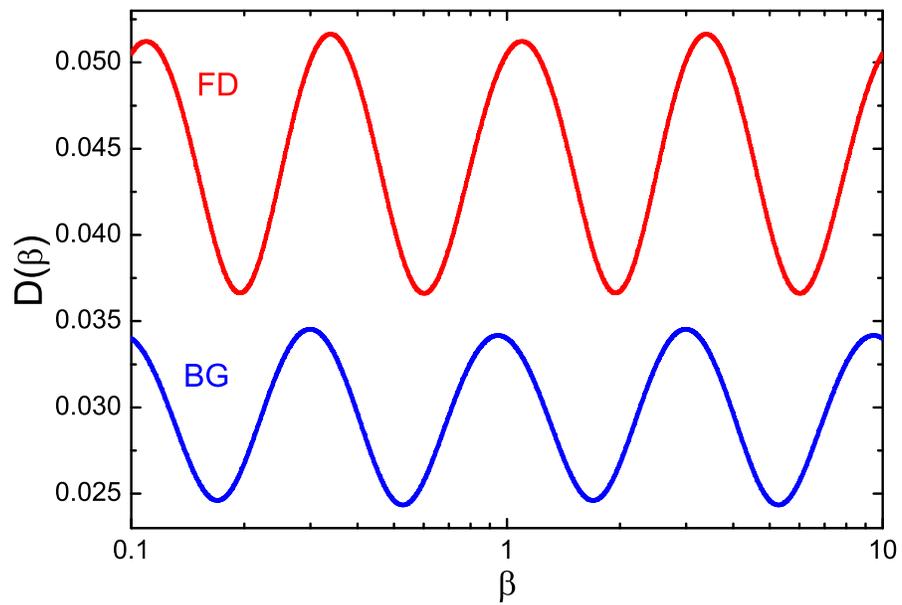}
\caption{The distances of the first digit distributions of BG and FD
statistics from Benford's law versus the inverse temperature $\beta
= 1/kT$.\label{pd}}
\end{figure}

Here we do not hope to mislead readers, actually, the variations
from Benford's law are always beneath the illustrated maximum
deviations, and the maximum deviations never appear simultaneously
for different digits. The total distance $D(\beta)$ defined as
\begin{equation}
D(\beta) = \sqrt{\sum_{d=1}^9 \left[ P(d;\beta)-P_{\rm Ben}(d) \right]^2}
\end{equation}
for BG and FD statistics versus temperature is depicted in
Fig.~\ref{pd}. They both have a wavy shape with the same periodicity
as the first digit distributions. We can see again that $D(\beta)$
of BG statistics, within the bound of 0.035, is always smaller than
that of FD statistics, which is in the range of 0.035--0.055.

\begin{figure}
\includegraphics[width=13.5cm]{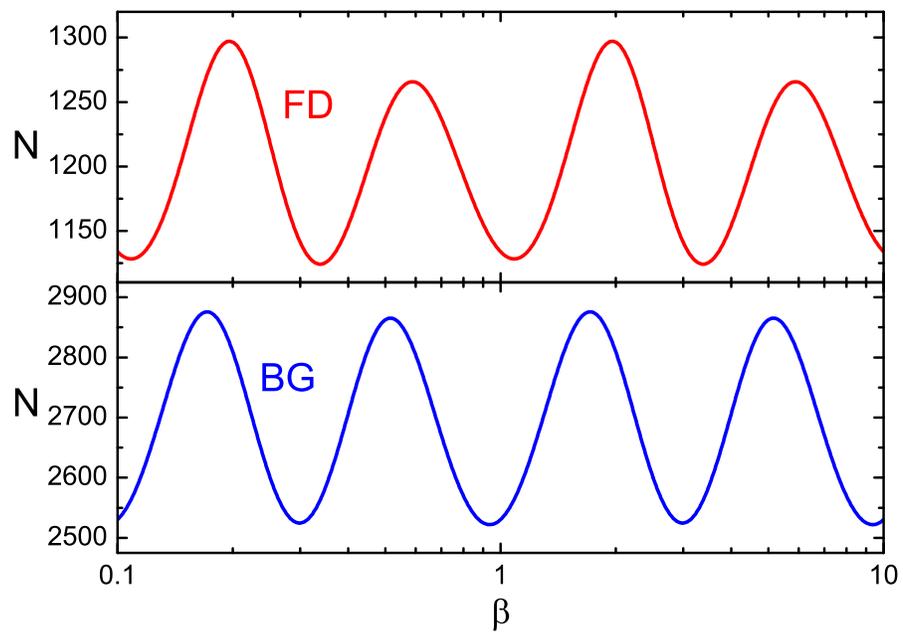}
\caption{The minimum sizes of samples required to distinguish the
first digit distributions of  BG (lower panel) and FD (upper panel)
statistics from Benford's law versus the inverse temperature $\beta
= 1/kT$, under the confidence level 95\%.\label{pn}}
\end{figure}

As shown and discussed above (cf. Fig.~\ref{perr} and
Fig.~\ref{pd}), from the measured data sets based on the BG or FD
statistics, the variations are rather small, even not significant
enough to distinguish from the Benford distribution if the sample is
not sufficiently large. When estimating the fitness of the observed
probability distribution $P_{\rm obs}(d)$ to the theoretical one
$P_{\rm Ben}(d)$, we use fitness estimating $\chi^2$, namely Pearson
$\chi^2$,
\begin{equation}\label{chisq}
\chi^2(n-1) = \sum_{d=1}^n N \times
\frac{\left[ P_{\rm obs}(d) - P_{\rm Ben}(d) \right]^2}{P_{\rm Ben}(d)}
\end{equation}
where $N$ is the size of the sample and here in our question $n=9$.
In Eq.~(\ref{chisq}), the degree of freedom is $9 - 1 = 8$, and
under the confidence level 95\%, $\chi^2(8) = 15.507$. The minimum
sizes of the samples required to distinguish the first digit
distributions of BG and FD statistics from Benford's law are
illustrated in Fig.~\ref{pn}, under the confidence level 95\% and
the assumption that the sample is virtually distributed exactly
following Eq.~(\ref{bg}) and Eq.~(\ref{fd}). Since the BG
distribution follows Benford's law more closely than the FD
distribution, its minimum size of the sample required to make a
distinction from the Benford distribution is larger thereof, more
than two times of that for the FD distribution. However, the desired
samples are more than one or several thousands for both cases to
tell the difference, and even larger when the criterion is made more
strict, e.g., under the confidence level 90\%. Because of the
limitation to collect data, it appears somehow unavailable under
most conditions to judge between the Benford distribution and
physical statistics at a given temperature.

Moreover, due to the variation of temperature $T$ and hence $\beta$
(cf. superstatistics~\cite{bc03} for an example), the difference is
further smoothed down. Then, in real world, the distributions
strictly obeying the BG statistics or FD statistics are supposed to
appear as the Benford distribution within rather good precision or
in an exactly way.

We now return to the 20 different tables consisting of over 20,000
entries investigated extensively by Benford~\cite{b38}. Though
Diaconis and Freedman provided evidence that Benford had manipulated
round-off errors~\cite{df79}, the raw data are also a remarkably
good fit. However, Raimi pointed out that, ``what came closest of
all, however, was the union of all his tables''~\cite{r69b}. The
combination of data from various unrelated domains can give a
marvelously perfect fit to Benford's law. An important breakthrough
motivated by it was achieved by Hill in 1995. He proved that ``if
distributions are selected at random (in any `unbiased' way) and
random samples are then taken from each of these distributions, the
significant digits of the combined sample will converge to the
logarithmic (Benford) distribution''~\cite{h95c}. As for our study,
the first digits of different samples picked from different systems
at different temperatures are expected to converge to the
intermediate value or the integral mean, $P(d) = \log_{10}(1+1/d)$,
which is exactly Benford's law.

The inclusion of chemical potential $\mu$ has some influence on the
numerical results, except for the BG distribution where the effect
is exactly canceled out by the normalization requirement. Here in
our paper, we only preliminarily considered the case $\mu=0$ for
simplicity. For the non-vanishing $\mu$, the chemical potential can
depend on the temperature, and a concrete calculation is needed. As
a simple case study, we adopt
\begin{equation}
f(E) \propto \frac{1}{e^{\beta E} \pm z}
\end{equation}
where plus and minus signs correspond to FD and BE distributions
respectively, and $z=e^{\beta \mu}$ is the fugacity. Then
$P(d;\beta,z)$ becomes
\begin{equation}
P(d; \beta,z) = \sum_{n = -\infty}^{\infty} \frac{1}{\ln(1\pm z)}
\left\{ \beta \cdot 10^n + \ln \left[ \frac{e^{\beta d \cdot
10^n} \pm z}{e^{\beta (d+1) \cdot 10^n} \pm z} \right] \right\},
\end{equation}
which is very similar to Eq.~(\ref{fdpd}). Relevant properties
discussed before, as well as the conformance to Benford's law, are
all preserved. Moreover, for the BE distribution where $\mu$ can
only take a non-positive value, hence $z\leq 1$, the distribution
converges accordingly.

The last points we stress are the two most important features of
Benford's law which contribute to make it so famous, namely,
scale-invariance and base-invariance. For scale-invariance, the
change of unit of the energy $E$ in our study equals to rescale the
temperature $T$ or the inverse temperature $\beta$. As mentioned,
the multiplicative appearance of $\beta$ and $E$ leads to the
periodic property with respect to the logarithm of $\beta$. Thus the
scale transformation might equivalently change $\beta$, but not the
global probability distribution, especially, the intermediate value
and the integral mean. As for base-invariance, we stress that the
analysis in this paper does not depend on the base $b$ very much.
However, as noted in Ref.~\cite{el03} for exponential random
variables (in our study, corresponding to the BG distribution),
larger $b$ will result in larger variations.

\section{Summary}

Statistical physics underpins the concept of ergodicity, which means that all accessible
microstates are equiprobable over a long period of time, and
the cooperative effects and nonlinear dynamics are important to lead to sufficient statistics.
On the other hand, a peculiar digit law, named Benford's law, concerns
the digit statistics in various domains, and elegantly presents the
complicated dynamics and
global regularities of the nature in its compact formula.

We relate Benford's law with three kinds of extensively used
physical statistics, i.e., the Boltzmann-Gibbs statistics, the
Fermi-Dirac statistics, and the Bose-Einstein statistics. It is
found that all three distributions fluctuate slightly around or
exactly conform to the Benford distribution, and their intermediate
values and the integral means converge to Benford's law exactly. It
seems that the Benford distribution is a general pattern in
statistical physics. Thus it turns out that Benford's law might be a
more profound and fundamental law than those in physical statistics,
especially in the fields where thermal statistics, even nonthermal
statistics, is invalidated, where Benford's law still applies well.
Moreover, the details of the mantissa distribution are also
discussed, and we suggest that the inverse distribution of mantissa
might present an important clue to look into deeper reasons and
crucial aspects of the logarithmic digit law.

For the moment, the Benford's law has been studied mainly
mathematically, but not so much physically. We need to understand
more physical significance of this law, and a central concerning is
when it works and when it does not, and the underlying reason why it
works. Further researches are needed to understand more aspects of
this significant digit law, which has revealed a mysterious
regularity in the realistic world and remains elusive for more than
one hundred years.

\section*{Acknowledgments}

This work is partially supported by National Natural Science
Foundation of China (Nos.~10721063, 10975003). It is also supported
by Hui-Chun Chin and Tsung-Dao Lee Chinese Undergraduate Research
Endowment (Chun-Tsung Endowment) at Peking University, and by
National Fund for Fostering Talents of Basic Science (Nos.~J0630311,
J0730316).

\bibliographystyle{elsarticle-num}

\clearpage

{\fontsize{18pt}{\baselineskip}\selectfont Figure captions}
\begin{itemize}
\item Figure 1:~~~The comparisons of the first digit distributions
for BG distribution and Benford's law via the inverse of temperature
$\beta = 1/kT$, where $k$ is the Boltzmann constant. The horizontal
line in each panel presents the value predicted by Benford's law
(also see Figure 1 in Ref.~\cite{el03}).

\item Figure 2:~~~The comparisons of the first digit distributions
for FD distribution and Benford's law via the inverse of temperature
$\beta = 1/kT$, where $k$ is the Boltzmann constant. The horizontal
line in each panel presents the value predicted by Benford's law.

\item Figure 3:~~~{\it Upper panel:} The maximum deviation of the first digit
distributions from Benford's law for the BG distribution (squared)
and the FD distribution (star-shaped); {\it Lower panel:} the
maximum relative deviation of the first digit distributions from
Benford's law for the BG distribution (squared) and FD distribution
(star-shaped).

\item Figure 4:~~~The distances of the first digit distributions of
BG and FD statistics from Benford's law versus the inverse
temperature $\beta = 1/kT$.

\item Figure 5:~~~The minimum sizes of samples required to distinguish the
first digit distributions of BG (lower panel) and FD (upper panel)
statistics from Benford's law versus the inverse temperature $\beta
= 1/kT$, under the confidence level 95\%.
\end{itemize}

\clearpage

\begin{table}
\begin{center}
\caption{The maximum deviations of the first digit distributions
from Benford's law for the BG distribution and the FD
distribution.}\vspace{0.5cm} {\begin{tabular}{cccccccc}
\hline %
\hline %
$d$ & $P_{\rm Ben}(d)$ &~~& \multicolumn{2}{c}{Max $|\Delta P(d)|$}
&~~&
\multicolumn{2}{c}{Max $|\Delta P(d) / P_{\rm Ben}(d)|$} \\
First Digit & Benford &~~& BG & FD &~~& BG & FD\\
\hline %
1 & 0.301 &~~& 0.0291 & 0.0435 &~~& 9.66\% & 14.5\% \\
2 & 0.176 &~~& 0.0187 & 0.0281 &~~& 10.6\% & 15.9\% \\
3 & 0.125 &~~& 0.0136 & 0.0204 &~~& 10.9\% & 16.3\% \\
4 & 0.097 &~~& 0.0107 & 0.0160 &~~& 11.0\% & 16.5\% \\
5 & 0.079 &~~& 0.0088 & 0.0131 &~~& 11.1\% & 16.6\% \\
6 & 0.067 &~~& 0.0074 & 0.0111 &~~& 11.1\% & 16.6\% \\
7 & 0.058 &~~& 0.0064 & 0.0097 &~~& 11.1\% & 16.7\% \\
8 & 0.051 &~~& 0.0057 & 0.0085 &~~& 11.1\% & 16.7\% \\
9 & 0.046 &~~& 0.0051 & 0.0076 &~~& 11.1\% & 16.7\% \\
\hline %
\end{tabular}\label{terr}}
\end{center}
\end{table}

\clearpage

\clearpage

\clearpage

\clearpage

\clearpage

\end{document}